\documentclass[
preprint, pra,showpacs,nofootinbib]{revtex4}
\usepackage{float,epsfig}
\usepackage{bm}% bold math
\usepackage{graphicx}% Include figure files
\usepackage{subfigure}
\usepackage{dcolumn}% Align table columns on decimal point
\usepackage{amsmath,amssymb,amsthm}
\usepackage{bm}
\def\/{\over}
\begin{document}
\title{\bf Modification of energy shifts of atoms by the presence of a boundary in a thermal bath and the Casimir-Polder force}
\author{ Zhiying Zhu and  Hongwei Yu\footnote{Corresponding author} }
%\email{hwyu@cosmos.phy.tufts.edu}
\affiliation{
Department of Physics and Institute of  Physics,\\
Hunan Normal University, Changsha, Hunan 410081, China\\
Key Laboratory of Low Dimensional Quantum Structures and Quantum
Control of Ministry of Education, Hunan Normal University ,
Changsha, Hunan 410081, China
 }

%\date{\today}

\begin{abstract}
We study the modification by the presence of a plane wall of energy
level shifts of two-level atoms which are in multipolar coupling
with quantized electromagnetic fields in a thermal bath in a
formalism which separates the contributions of thermal fluctuations
and radiation reaction and allows a distinct treatment to atoms in
the ground and excited states. The position dependent energy shifts
give rise to an induced force acting on the atoms.  We are able to
identify three different regimes where the force shows distinct
features and examine, in all regimes, the behaviors of this force in
both the low temperature limit and the high temperature limit for
both the ground state and excited state atoms, thus providing some
physical insights into the atom-wall interaction at finite
temperature. In particular, we show that both the magnitude and the
direction of the force acting on an atom may have a clear dependence
on atomic the polarization directions. In certain cases, a change of
relative ratio of polarizations in different directions may result
in a change of direction of the force.
\end{abstract}


\maketitle

\section{Introduction}

Quantum fields constrained by the presence of boundaries exhibit
many interesting properties because the presence of boundaries
disturbs the modes of quantum fields. As a result,  fluctuations of
quantum fields are altered, giving rise to a lot of novel effects,
such as the Casimir~\cite{Casimir} and Casimir-Polder
forces~\cite{CP}, the light-cone fluctuations when gravity is
quantized~\cite{Yu1,Yu2,Yu3}, the Brownian (random) motion of test
particles in an electromagnetic
vacuum~\cite{Yu4,Yu5,Tan,Yu7,SWu,HWL}, and modifications of
radiative properties of atoms in cavities such as the natural
lifetimes and energy level shifts which have begun to be
demonstrated in experiment~\cite{Brune,Marrocco}.

In the present paper, we are particularly interested in the
Casmir-Polder force at finite temperature. The Casimir force, which
has now been measured with precision~\cite{SBDSH93,DD03}, is not
only fascinating in theoretical research but also increasingly
important in technological applications, such as
nanophysics~\cite{Craighead00}, chemical identification of surface
atoms via atomic force microscope~\cite{Sugimoto07} and the
construction of novel biomimetic dry adhesives~\cite{NT05}.

Casimir-Polder force at zero temperature is usually regarded as a
result of the reshaping of vacuum fluctuations induced by the
presence of boundaries. In a thorough understanding of the
Casimir-Polder force one must also consider the effects of thermal
fluctuations in addition to the vacuum ones. It should be pointed
out that in the studies of thermal Casimir-Polder force, a majority
of researches invokes a macroscopic approach to atoms pioneered by
Lifshitz \cite{Lifshitz,Antezza04,Babb04} or a linear response
theory of atoms \cite{Antezza05,McLachlan63}, although some uses a
QED treatment of the atom-field coupling
\cite{Nakajima97,Gorza06,Buhmann08}.

With an atom-field coupling, the Casimir-Polder force could be
understood as a result of the position dependent modification of
radiative properties of atoms (the energy level shifts in
particular) in the confined space in QED quantitatively. Let us
note, however, that a clear and simple physical interpretation does
not always emerge from the QED description. One interpretation is
that the boundary induced effects on atom's radiative properties may
be explained by the modification of vacuum fluctuations of the
quantum electromagnetic field
\cite{Welton48,Agarwal75,Spruch78,Ravndal83}. On the other hand, the
same effect can also be described as a consequence of the reaction
of the instantaneous atomic dipole to its own radiation field
reflected from the boundary
\cite{Wylie84,Barut87,Boyer80,Milonni73}. In fact, it has been shown
within QED that the extent to which each mechanism contributes to
the total effect can be chosen arbitrarily just by changing the
ordering of the atom and field operators in the interaction
Hamiltonian\cite{Ackerhalt73,Senitzky73,Smith73,Milonni76}.

As a result, there exists an indetermination in the separation of
effects of vacuum fluctuations and radiation reaction such that
distinct contributions of vacuum fluctuations and radiation reaction
to the radiative properties of atoms do not possess an independent
physical meaning. Therefore, although quantitative results for
spontaneous emission and energy level shifts are well-established,
the physical interpretations remained controversial until Dalibard,
Dupont-Roc and Cohen-Tannoudji(DDC) argued in \cite{Dalibard82} and
\cite{Dalibard84} that there exists a symmetric operator ordering of
atom and field variables where the distinct contributions of vacuum
fluctuations and radiation reaction to the rate of  change of an
atomic observable are separately Hermitian.
%If one demands such an
%ordering, an independent physical meaning can be assigned to each
%contribution.
Recently, this formalism has been employed to study the radiative
properties of atoms in various cases
\cite{Meschede90,Audretsch94,Audretsch95,Audretsch95CQL,Passante98,Yu8,Zhi,Yu9,Yu10,Yu11,Rizzuto,Zhi2}.
In the present paper, we would like to use the DDC formalism to
calculate the modification of energy shifts of atoms by the presence
of a boundary in a thermal bath and the forces thus induced between
the atom and the boundary. This is not a new problem. However, by
using the DDC formalism which separates the contributions of thermal
fluctuations (including vacuum fluctuations) and radiation reaction
and allows a distinct microscopic treatment to atoms in the ground
and excited states, in contrast to the macroscopic approach where
atoms are treated as a limiting case of a dielectric, we hope to
gain some new physical insight into the atom-wall interactions. Let
us note that the same problem at zero temperature has already been
discussed in using the DDC formalism~\cite{Meschede90}.

The paper is organized as follows, we give, in Sec.~II, a review of
the general formalism developed in Ref.~\cite{Dalibard84} and
generalized in Ref.~\cite{Audretsch95,Passante98} to the case of a
neutral polarizable two-level atom interacting with  quantized
electromagnetic fields at finite temperature $T$, and apply it, in
Sec.~III, to calculate the position dependent energy shifts, which
give rise to an induced force acting on the atom. Then we will
study, in detail, this force in both the low temperature limit and
the high temperature limit for both the ground state and excited
state atoms. Finally, we will conclude in Sec.~IV with a summary of
results obtained.

\section{the thermal fluctuations and radiation reaction contributions}

We consider a neutral polarizable two-level atom in interaction with
 quantized electromagnetic fields in a thermal bath of temperature $T$ in the
presence of an infinite conducting plane wall located at $z=0$. We
assume that the atom is at rest at a distance, $z$, from the wall,
and has stationary states $|-\rangle$ and $|+\rangle$, with energies
$-{1\/2}\hbar\omega_0$ and $+{1\/2}\hbar\omega_0$ and a level
spacing $\hbar\omega_0$.

To investigate the modification of the atom's energy shifts by the
presence of the plane boundary, we will first review the general
formalism developed by DDC which allows a distinct separation of the
contributions of thermal fluctuations (including  zero temperature
vacuum fluctuations) and radiation reaction to the energy shifts of
atomic levels.

The atom is supposed to interact with the quantized field in the
multipolar coupling scheme \cite{CPP95}, so that the Hamiltonian
that describes the atom-field system with respect to the proper time
$\tau$ can be written as
\begin{eqnarray}
H(\tau)=H_A(\tau)+H_F(\tau)+H_I(\tau)\;,\label{H}
\end{eqnarray}
where
\begin{eqnarray}
H_A(\tau)=\hbar\omega_0S_z(\tau)\;,
\end{eqnarray}
\begin{eqnarray}
H_F(\tau)=\sum_k \hbar\omega_{\vec{k}} a_{\vec{k}}^\dag a_{\vec{k
}}{dt\/d \tau} \;,
\end{eqnarray}
\begin{eqnarray}
H_I(\tau)=-\bm{\mu}(\tau)\cdot \textbf{E}(x(\tau))=-\bm{\mu}\cdot
\textbf{E}(x(\tau))[S_+(\tau)+S_-(\tau)]\;.\label{HI}
\end{eqnarray}
Here $\vec{k}$ denotes the wave vector and polarization of the field
modes, $\bm{\mu}$ the atomic electric dipole moment, and the
electric field operator, \textbf{E}, is evaluated along the
trajectory $x(\tau)$ of the atom. $S_z$, $S_+$, $S_-$ are the
pseudospin operators of the two-level atom, and
$S_z(0)={1\/2}|+\rangle\langle+|-{1\/2}|-\rangle\langle-|$,
$S_+(0)=|+\rangle\langle-|$, $S_-(0)=|-\rangle\langle+|$.

With the Hamiltonian of the system given, we can write down the
Heisenberg equations of motion for the dynamical variables of the
atom and field from Eq.~(\ref{H}). The solutions of the equations of
motion can be split into the two parts: a free part, which is
present even in the absence of the coupling, and a source part,
which is induced by the interaction of the atom and the field. We
assume that the electromagnetic field is in a thermal state with the
density matrix $\rho=e^{-\beta H_F/(\hbar c)}$, where $\beta={\hbar
c/(Tk_B)}$ is the thermal wavelength, and the atom is initially in
the state $|b\rangle$. $k_B$ is the Boltzmann constant. To identify
the contributions of thermal fluctuations and radiation reaction to
the radiative atomic energy shifts, we choose, following DDC,  a
symmetric ordering between the atom and the field variables and
consider the effects of $E^f$ (corresponding to the effect of
thermal fluctuations) and $E^s$ (corresponding to the effect of
radiation reaction) separately in the Heisenberg equations of an
arbitrary atomic observable $G$. Following the procedures that have
been shown in Refs.~\cite{Dalibard84,Audretsch95,Passante98}, we
take the thermal average over the electromagnetic field and
decompose various rates of $G$ into a Hamiltonian part, describing
the energy levels of the atom shifted due to coupling with the
electromagnetic field, and a non-Hamiltonian part, describing the
effects of relaxation, then in a perturbation treatment up to order
$\mu^2$, we find
\begin{eqnarray}
\bigg\langle\beta\bigg|\bigg({d
G(\tau)\/d\tau}\bigg)_{tf,rr}\bigg|\beta\bigg\rangle={i\/\hbar}[H^{eff}_{tf,rr}(\tau),G(\tau)]+\text{non-Hamiltonian\
terms}\;,
\end{eqnarray}
where
\begin{eqnarray}
H^{eff}_{tf}(\tau)=-{i\/2\hbar}\int^\tau_{\tau_0}d\tau'(C^F_{ij})_\beta(x(\tau),x(\tau'))[\mu_i^f(\tau),\mu_j^f(\tau')]\;,\label{Hvf}
\end{eqnarray}
\begin{eqnarray}
H^{eff}_{rr}(\tau)=-{i\/2\hbar}\int^\tau_{\tau_0}d\tau'(\chi^F_{ij})_\beta(x(\tau),x(\tau'))\{\mu_i^f(\tau),\mu_j^f(\tau')\}\;.\label{Hrr}
\end{eqnarray}
Here $[\ ,\ ]$ and $\{\ ,\ \}$  denote the commutator and
anticommutator respectively while subscript `` $tf$ " stands for
thermal fluctuations and ``$rr$" for radiation reaction. The
statistical functions $(C^F_{ij})_\beta$ and $(\chi^F_{ij})_\beta$
are defined as
\begin{eqnarray}
(C^F_{ij})_\beta(x(\tau),x(\tau'))={1\/2}\langle\{E_i^f(x(\tau)),E_j^f(x(\tau'))\}\rangle_\beta\;,\label{cf}
\end{eqnarray}
\begin{eqnarray}
(\chi^F_{ij})_\beta(x(\tau),x(\tau'))={1\/2}\langle[E_i^f(x(\tau)),E_j^f(x(\tau'))]\rangle_\beta\;,\label{xf}
\end{eqnarray}
which are also called the symmetric correlation function and the
linear susceptibility of the field. Taking the expectation value of
Eqs.~(\ref{Hvf}) and (\ref{Hrr}) in the atom's initial state
$|b\rangle$, we can obtain the contributions of thermal fluctuations
and radiation reaction to the energy shifts of the atom's level
$|b\rangle$,
\begin{eqnarray}
(\delta
E_b)_{tf}=-{i\/\hbar}\int^\tau_{\tau_0}d\tau'(C^F_{ij})_\beta(x(\tau),x(\tau'))(\chi^A_{ij})_b(\tau,\tau')\;,\label{Evf}
\end{eqnarray}
\begin{eqnarray}
(\delta
E_b)_{rr}=-{i\/\hbar}\int^\tau_{\tau_0}d\tau'(\chi^F_{ij})_\beta(x(\tau),x(\tau'))(C^A_{ij})_b(\tau,\tau')\;,\label{Err}
\end{eqnarray}
where  $(C^A_{ij})_b$ and $(\chi^A_{ij})_b$, the symmetric
correlation function and the linear susceptibility of the atom, are
defined as
\begin{eqnarray}
(C^A_{ij})_b(\tau,\tau')={1\/2}\langle
b|\{\mu_i^f(\tau),\mu_j^f(\tau')\}|b\rangle\;,
\end{eqnarray}
\begin{eqnarray}
(\chi^A_{ij})_b(\tau,\tau')={1\/2}\langle
b|[\mu_i^f(\tau),\mu_j^f(\tau')]|b\rangle\;,
\end{eqnarray}
which are characterized by the atom itself. Explicitly, the
statistical functions of the atom can be given as
\begin{eqnarray}
(C_{ij}^A)_b(\tau,\tau')={1\/2}\sum_d[\langle
b|\mu_i(0)|d\rangle\langle d|\mu_j(0)|b\rangle
e^{i\omega_{bd}(\tau-\tau')}+\langle b|\mu_j(0)|d\rangle\langle
d|\mu_i(0)|b\rangle e^{-i\omega_{bd}(\tau-\tau')}]\label{ca}\;,\nonumber\\
\end{eqnarray}\begin{eqnarray}
(\chi_{ij}^A)_b(\tau,\tau')={1\/2}\sum_d[\langle
b|\mu_i(0)|d\rangle\langle d|\mu_j(0)|b\rangle
e^{i\omega_{bd}(\tau-\tau')}-\langle b|\mu_j(0)|d\rangle\langle
d|\mu_i(0)|b\rangle e^{-i\omega_{bd}(\tau-\tau')}]\label{xa}\;,\nonumber\\
\end{eqnarray}
where $\omega_{bd}=\omega_b-\omega_d$ and the sum extends over a
complete set of atomic states. In order to calculate the statistical
functions of the electric field, Eqs.~(\ref{cf}) and (\ref{xf}), we
will use the two point function of the four potential, $A^\mu(x)$,
at finite temperature, which can be obtained  by the method of
images both in imaginary-time and in space. In the presence of a
boundary, one finds in the Feynman gauge ~\cite{Yu7},
\begin{eqnarray}
D^{\mu\nu}_\beta(x,x')=\langle
A^\mu(x)A^\nu(x')\rangle_\beta=D_{\beta,free}^{\mu\nu}(x-x')
+D_{\beta,bnd}^{\mu\nu}(x,x')\label{D}\;,
\end{eqnarray}
where
\begin{eqnarray}
D_{\beta,free}^{\mu\nu}(x-x')={\hbar\/4\pi^2\varepsilon_0c}\sum_{k=-\infty}^\infty{\eta^{\mu\nu}\/{
[(tc-t^\prime c+ik\beta-i\varepsilon)^2-(x-x^\prime)^2-(y-y^\prime)^2-(z-z^\prime)^2]}}\;,\nonumber\\
\end{eqnarray}
is the two point function in the free space and
\begin{eqnarray}
D_{\beta,bnd}^{\mu\nu}(x,x')=-{\hbar\/4\pi^2\varepsilon_0c}\sum_{k=-\infty}^\infty{{\eta^{\mu\nu}+2n^\mu
n^\nu}\/{ [(tc-t^\prime
c+ik\beta-i\varepsilon)^2-(x-x^\prime)^2-(y-y^\prime)^2-(z+z^\prime)^2]}}\;.\label{Dbound}\nonumber\\
\end{eqnarray}
represents the correction induced by the presence of the boundary.
Here $\varepsilon_0$ is the vacuum dielectric constant, the
subscript `` $bnd$ " stands for the part induced by the presence of
a boundary, $\varepsilon\rightarrow+0$,
$\eta^{\mu\nu}$=diag$(1,-1,-1,-1)$ and the unit normal vector
$n^\mu=(0,0,0,1)$.  Then, we can write the electric field two point
function as
\begin{eqnarray}
\langle E_i(x(\tau))E_j(x(\tau'))\rangle_\beta=\langle
E_i(x(\tau))E_j(x(\tau'))\rangle_{\beta,free} +\langle
E_i(x(\tau))E_j(x(\tau'))\rangle_{\beta,bnd}\;,\label{EE}
\end{eqnarray}
where
\begin{eqnarray}
\langle E_i(x(\tau))E_j(x(\tau'))\rangle_{\beta,free}&=&{\hbar
c\/4\pi^2\varepsilon_0}\sum_{k=-\infty}^\infty(\delta_{ij}\partial
_0\partial_0^\prime-\partial_i\partial_j^\prime)\nonumber\\&&\times
{1\/(x-x')^2+(y-y')^2+(z-z')^2-(tc -t'c+ik\beta-i\varepsilon)^2}\;,\nonumber\\
\end{eqnarray}
and
\begin{eqnarray}
\langle E_i(x(\tau))E_j(x(\tau'))\rangle_{\beta,bnd}&=&-{\hbar
c\/4\pi^2\varepsilon_0}\sum_{k=-\infty}^\infty[\,(\delta_{ij}-2n_in_j)\,\partial
_0\partial_0^\prime-\partial_i\partial_j^\prime\,]\nonumber\\&&\times
{1\/(x-x')^2+(y-y')^2+(z+z')^2-(tc-t'c+ik\beta-i\varepsilon)^2}\;.\label{EEb}\nonumber\\
\end{eqnarray}
Here $\partial^\prime_j$ denotes the differentiation with respect to
$x^\prime_j$. So, the statistical functions of the electric field
can be obtained from Eq.~(\ref{EE}) as a sum of the free space part
and the boundary-dependent part.

Our aim in this paper is to study the modification of atomic energy
shifts by the presence of the conducting plane boundary in a thermal
bath and the force thus induced on the atom. So in Eqs.~(\ref{Evf})
and (\ref{Err}), the contributions of thermal fluctuations and
radiation reaction to the energy shifts of the atom's level, we are
only interested in the boundary-dependent part,
\begin{eqnarray}
(\delta
E_b)_{tf}^{bnd}=-{i\/\hbar}\int^\tau_{\tau_0}d\tau'(C^F_{ij})_{\beta,bnd}(x(\tau),x(\tau'))(\chi^A_{ij})_b(\tau,\tau')\;,\label{Ecpvf}
\end{eqnarray}
\begin{eqnarray}
(\delta
E_b)_{rr}^{bnd}=-{i\/\hbar}\int^\tau_{\tau_0}d\tau'(\chi^F_{ij})_{\beta,bnd}(x(\tau),x(\tau'))(C^A_{ij})_b(\tau,\tau')\;.\label{Ecprr}
\end{eqnarray}
Therefore the modification of energy shifts by the presence of the
conducting plane boundary in a thermal bath is given by
\begin{eqnarray}
(\delta E_b)_{tot}^{bnd}=(\delta E_b)_{tf}^{bnd}+(\delta
E_b)_{rr}^{bnd}\;.
\end{eqnarray}

\section{Modification of energy shifts by the presence of a boundary in a thermal bath}

In this section we will apply the formalism developed in the
previous section to calculate separately the contributions of the
thermal fluctuations and radiation reaction to the energy level
shifts of a static two-level atom induced by the presence of a
boundary in a thermal bath and the force acting on the atom as a
result. We first give the basic results and then proceed to detailed
discussions of the behaviors of the position-dependent energy shifts
and thus force on the atom  both in the low temperature limit (when
the wavelength of the thermal photons is much larger than the
transition wavelength of the atom) and the high temperature limit
(when the wavelength of the thermal photons is much smaller than the
transition wavelength of the atom).

\subsection{Basic results}
For an atom at rest, we can compute the boundary-dependent two point
function of the electric field in a thermal bath from the general
form Eq.~(\ref{EEb}) to get
\begin{eqnarray}
\langle E_i(x(\tau))E_j(x(\tau'))\rangle_{\beta,bnd}=-{\hbar
c\/\pi^2\varepsilon_0}
\sum_{k=-\infty}^\infty{(\delta_{ij}-2n_in_j)(uc+ik\beta)^2+4z^2\/[(uc+ik\beta-i\varepsilon)^2-4z^2]^3}\;,
\end{eqnarray}
where $u=\tau-\tau'$. Consequently, the boundary-dependent
statistical functions of the electric field can be expressed as
\begin{eqnarray}
(C_{ij}^F)_{\beta,bnd}(x(\tau),x(\tau'))=-{\hbar
c\/2\pi^2\varepsilon_0}\sum_{k=-\infty}^\infty&\bigg(&
{(\delta_{ij}-2n_in_j)(uc+ik\beta)^2+4z^2\/[(uc+ik\beta-i\varepsilon)^2-4z^2]^3}\nonumber\\&&+
{(\delta_{ij}-2n_in_j)(uc+ik\beta)^2+4z^2\/[(uc+ik\beta+i\varepsilon)^2-4z^2]^3}\bigg)\;,\label{cfinertial}
\end{eqnarray}
\begin{eqnarray}
(\chi_{ij}^F)_{\beta,bnd}(x(\tau),x(\tau'))=-{\hbar
c\/2\pi^2\varepsilon_0}\bigg(
{(\delta_{ij}-2n_in_j)u^2c^2+4z^2\/[(uc-i\varepsilon)^2-4z^2]^3}-
{(\delta_{ij}-2n_in_j)u^2c^2+4z^2\/[(uc+i\varepsilon)^2-4z^2]^3}\bigg)\;.\nonumber\\\label{xfinertial}
\end{eqnarray}
The linear susceptibility of the electromagnetic field,
Eq.~(\ref{xfinertial}), is independent of the temperature. Let us
note that only the diagonal components ($i=j=x,y,z$) of the
statistical functions are nonzero, and the components parallel to
the conducting plane, i.e., the $xx$ and $yy$ components, are equal.
Substituting the nonzero statistical functions of the atom and the
electric field into the general formulas (\ref{Ecpvf}) and
(\ref{Ecprr}), we can obtain the boundary-dependent contributions of
thermal fluctuations and radiation reaction to the energy shifts of
atomic levels,
\begin{eqnarray}
(\delta
E_b)^{bnd}_{tf}&=&{ic\/4\pi^2\varepsilon_0}\sum_{k=-\infty}^\infty\sum_{j~d}|\langle
b|\mu_j(0)|d\rangle|^2\int_0^\infty
du(e^{i\omega_{bd}u}-e^{-i\omega_{bd}u})\nonumber\\&&\times\bigg(
{(1-2n_jn_j)(uc+ik\beta)^2+4z^2\/[(uc+ik\beta-i\varepsilon)^2-4z^2]^3}+
{(1-2n_jn_j)(uc+ik\beta)^2+4z^2\/[(uc+ik\beta+i\varepsilon)^2-4z^2]^3}\bigg)\;,\label{inercpvf}
\end{eqnarray}
and
\begin{eqnarray}
(\delta
E_b)^{bnd}_{rr}&=&{ic\/4\pi^2\varepsilon_0}\sum_{j~d}|\langle
b|\mu_j(0)|d\rangle|^2\int_0^\infty
du(e^{i\omega_{bd}u}+e^{-i\omega_{bd}u})\nonumber\\&&\times\bigg(
{(1-2n_jn_j)u^2c^2+4z^2\/[(uc-i\varepsilon)^2-4z^2]^3}-
{(1-2n_jn_j)u^2c^2+4z^2\/[(uc+i\varepsilon)^2-4z^2]^3}\bigg)\;.\label{inercprr}
\end{eqnarray}
Here we have extend the range of integration to infinity for
sufficiently long times $\tau-\tau_0$. The integration can be
evaluated using contour integrals with the help of the residue
theorem. With a definition of the atomic static scalar
polarizability
\begin{eqnarray}
\alpha_0=\sum_{j}\alpha_j=\sum_{j~d}{2|\langle
b|\mu_j(0)|d\rangle|^2\/3\omega_{0}\hbar}\;,\label{alpha}
\end{eqnarray}
the results can be cast into
\begin{eqnarray}
(\delta
E_+)^{bnd}_{tf}={3\hbar\omega_{0}\alpha_j\/128\pi\varepsilon_0}\bigg[\bigg(1+{2\/e^{\beta\omega_{0}/c}-1}\bigg)f_{j}(\omega_{0},z)
-g_{j}(\omega_{0},z,\beta)\bigg]\;,\label{inercpvf2e}
\end{eqnarray}
for the contribution of thermal fluctuations to the energy level
shift of the excited state,
\begin{eqnarray}
(\delta
E_-)^{bnd}_{tf}=-{3\hbar\omega_{0}\alpha_j\/128\pi\varepsilon_0}\bigg[\bigg(1+{2\/e^{\beta\omega_{0}/c}-1}\bigg)f_{j}(\omega_{0},z)
-g_{j}(\omega_{0},z,\beta)\bigg]\;,\label{inercpvf2g}
\end{eqnarray}
for that of the ground state, and
\begin{eqnarray}
(\delta E_+)^{bnd}_{rr}=(\delta
E_-)^{bnd}_{rr}={3\hbar\omega_{0}\alpha_j\/128\pi\varepsilon_0}f_{j}(\omega_{0},z)\;,\label{inercprr2}
\end{eqnarray}
for the contribution of radiation reaction to the energy shifts of
both the ground and excited states. Here we have defined
\begin{eqnarray}
f_{x}(\omega_{0},z)=f_{y}(\omega_{0},z)=
{4z^2\omega_{0}^2-c^2\/z^3c^2}\cos(2z\omega_{0}/c)-{2\omega_{0}\/z^2c}\sin(2z\omega_{0}/c)\;,\label{fx}
\end{eqnarray}
\begin{eqnarray}
f_{z}(\omega_{0},z)=-{2\/z^3}\cos(2z\omega_{0}/c)-{4\omega_{0}\/z^2c}\sin(2z\omega_{0}/c)\;,
\end{eqnarray}
\begin{eqnarray}
g_{x}(\omega_{0},z,\beta)=g_{y}(\omega_{0},z,\beta)={64c\/\pi}\sum_{k=-\infty}^\infty\int_0^\infty
du{(uc+k\beta)^2-4z^2\/[(uc+k\beta)^2+4z^2]^3}e^{-\omega_{0}u}\;,
\end{eqnarray}
\begin{eqnarray}
g_{z}(\omega_{0},z,\beta)=-{64c\/\pi}\sum_{k=-\infty}^\infty\int_0^\infty
du{1\/[(uc+k\beta)^2+4z^2]^2}e^{-\omega_{0}u}\;,\label{gz}
\end{eqnarray}
and summation over repeated indexes is implied.
%As the distance $z$
%increase, the functions $f_{j}(\omega_{0},z)$ and
%$g_{j}(\omega_{0},z,\beta)$ all decrease and give the retardation
%effects of Eqs.~(\ref{inercpvf2e})-(\ref{inercprr2}).
We note that the contribution of the radiation reaction to the
energy shifts, Eq.~(\ref{inercprr2}), does not depend on the
temperature  and the contribution of fluctuations of fields to the
energy level shift of the ground state is just opposite to that of
the excited state. Adding up the contributions of thermal
fluctuations and radiation reaction, we arrive at the
position-dependent energy shift,
\begin{eqnarray}
(\delta
E_+)^{bnd}_{tot}={3\hbar\omega_{0}\alpha_j\/128\pi\varepsilon_0}
\bigg[\bigg(2+{2\/e^{\beta\omega_{0}/c}-1}\bigg)f_{j}(\omega_{0},z)
-g_{j}(\omega_{0},z,\beta)\bigg]\;,\label{CPe}
\end{eqnarray}
for the excited state, and
\begin{eqnarray}
(\delta
E_-)^{bnd}_{tot}=-{3\hbar\omega_{0}\alpha_j\/128\pi\varepsilon_0}
\bigg[{2\/e^{\beta\omega_{0}/c}-1}f_{j}(\omega_{0},z)-g_{j}(\omega_{0},z,\beta)\bigg]\;,\label{CPg}
\end{eqnarray}
for the ground state. Since the energy shifts depend on the distance
$z$ between the atom and the conducting plane, the atom feels a
force near the wall. In what follows,  we examine in detail the
behaviors of this force in various circumstances, considering
separately the ground and excited atoms.

\subsection{Low temperature limit}

Let us now discuss the case when the wavelength of thermal photons
is much larger than the transition wavelength of the atom,
$\lambda_0$,
\begin{eqnarray}
\beta\gg\lambda_0\;,\label{low}
\end{eqnarray}
where $\lambda_0=c/\omega_0$. This condition, which makes
transitions from the ground state to the excited state virtually
impossible, is very well satisfied at the usual room temperature. As
a matter of fact, the first optical resonance of Rb atoms is as high
as $1.8\times 10^4$ K.  We will analyze how the interaction energy
or the force acting on the atom behaves as the distance varies in
three different regimes: the short distance, where the distance $z$
is much smaller than the transition wavelength of the atom
($z\ll\lambda_0$), the intermediate distance, where the distance $z$
is much larger than the transition wavelength of the atom but much
smaller than the thermal wavelength ($\lambda_0\ll z\ll\beta$), and
the long distance, where the distance $z$ is much larger than the
thermal wavelength ($z\gg\beta\gg\lambda_0$). For these three
distinct regimes, we will discuss the force on a ground state atom
and an excited one separately.

\subsubsection{ Ground state atom}

Let us start with the short distance regime, i.e., when
$z\ll\lambda_0$. In this case, thermal fluctuations and radiation
reaction contributions to the boundary-dependent energy shift become
approximately,
\begin{eqnarray}
(\delta
E_-)^{bnd}_{tf}\approx-{\hbar\/4\pi\varepsilon_0}&\bigg[&-{3\omega_0^2\alpha_z\/4\pi
cz^2}-{\omega_0^4\/c^3}(\alpha_x+\alpha_y-\alpha_z)\log
(z\omega_0/c)\nonumber\\&&-\bigg({2\pi^3c\/15\beta^4}+{16\pi^5c^3\/63\omega_{0}^2\beta^6}\bigg)(\alpha_x+\alpha_y-\alpha_z)
+{32\pi^5cz^2\/315\beta^6}(2\alpha_x+2\alpha_y-\alpha_z)\bigg]\;,\nonumber\\
\end{eqnarray}
and
\begin{eqnarray}
(\delta
E_-)^{bnd}_{rr}\approx-{\hbar\/4\pi\varepsilon_0}{3\omega_{0}\/32z^3}(\alpha_x+\alpha_y+2\alpha_z)\;.
\end{eqnarray}
Here the contribution of radiation reaction is much larger than that
of thermal fluctuations, and plays the dominant role in the total
energy shift. Consider for example, a Rb atom
($\omega_0=2.37\times10^{15}s^{-1}$) placed $1$ nanometer from the
conducting boundary at room temperature ($T=300K$), then the
contribution of radiation reaction is about $200$ times as large as
that of thermal fluctuations.  The closer the atom is to the
boundary, the larger is this relative ratio.

Since we are interested in the force on the atom, we will drop the
$z-$independent terms. Then we find by adding up the two
contributions
\begin{eqnarray} (\delta
E_-)^{bnd}_{tot}\approx-{\hbar\/4\pi\varepsilon_0}&\bigg[&{3\omega_{0}\/32z^3}(\alpha_x+\alpha_y+2\alpha_z)-{3\omega_0^2\alpha_z\/4\pi
cz^2}-{\omega_0^4\/c^3}(\alpha_x+\alpha_y)\log (z\omega_0/c)\nonumber\\
&&+{32\pi^5cz^2\/315\beta^6}(2\alpha_x+2\alpha_y-\alpha_z)\bigg]\;.\label{lowTlowz}
\end{eqnarray}
The first (leading) term on the righthand side of
Eq.~(\ref{lowTlowz}) is just the atom-wall interaction energy at
zero temperature in the short distance limit obtained in
Ref.~\cite{CP} (an extra factor of ${\hbar/(4\pi\varepsilon_0)}$  is
caused by the SI units used in the present paper) while the last
term gives leading thermal corrections. The leading position
dependent energy shift leads to the van der Waals force and the
calculations above show clearly that the source of this force is
purely radiation reaction. The energy shift given in
Eq.~(\ref{lowTlowz}) induces an attractive force between a ground
state atom and the plane wall. It is also interesting to note that
the atom's polarization in the longitudinal and transverse
directions are weighted differently in terms of their contribution
to the force. For a ground state atom polarized isotropically the
position dependent interaction energy shift reads
\begin{eqnarray}
(\delta
E_-)^{bnd}_{tot}\approx-{\hbar\/4\pi\varepsilon_0}\bigg({\omega_{0}\alpha_0\/8z^3}
+{32\pi^5c\alpha_0z^2\/315\beta^6}\bigg)\;,\label{lowTlowz2}
\end{eqnarray}
where the first term is just the usual van der Waals result. One
should bear in mind, however, that the polarization in different
directions contributes differently in the van der Waals force even
for isotropically polarized atoms. In fact, the van der Waals force
on a ground state atom polarizable only in $z$ direction would be
twice as that polarizable only in the parallel directions. This is
one of the new understandings we gain by not treating the atom as a
limiting case of a macroscopic dielectric.

\begin{figure}[htbp]\centering
\subfigure[]{\label{fig:inte:tfrr}
\includegraphics[height=2.3in,width=3.0in]{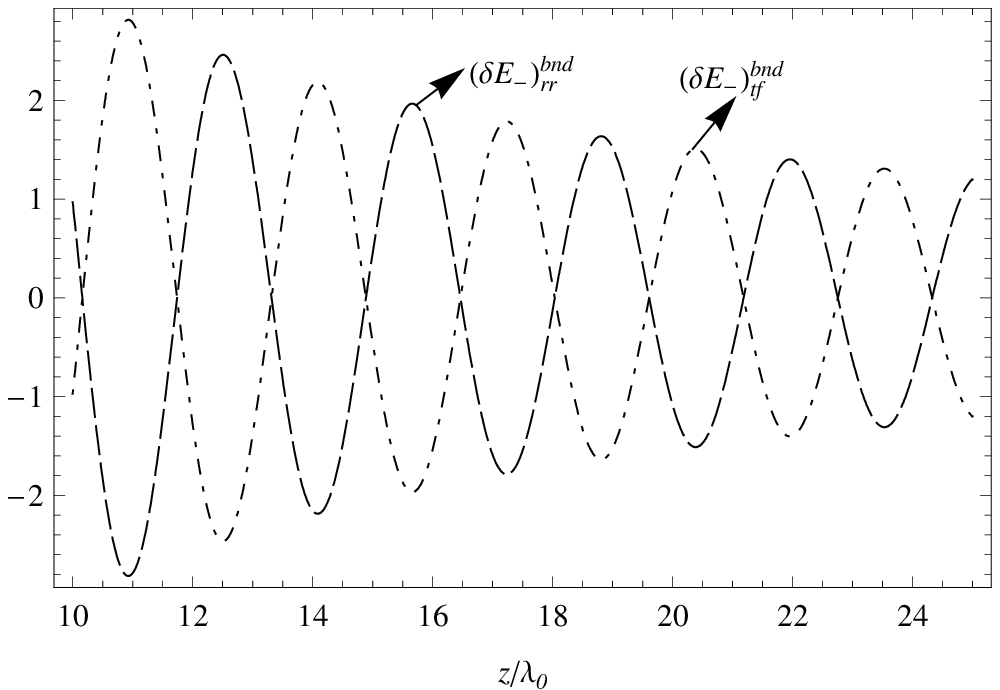}}
\subfigure[]{\label{fig:inte:tot}
\includegraphics[height=2.3in,width=3.0in]{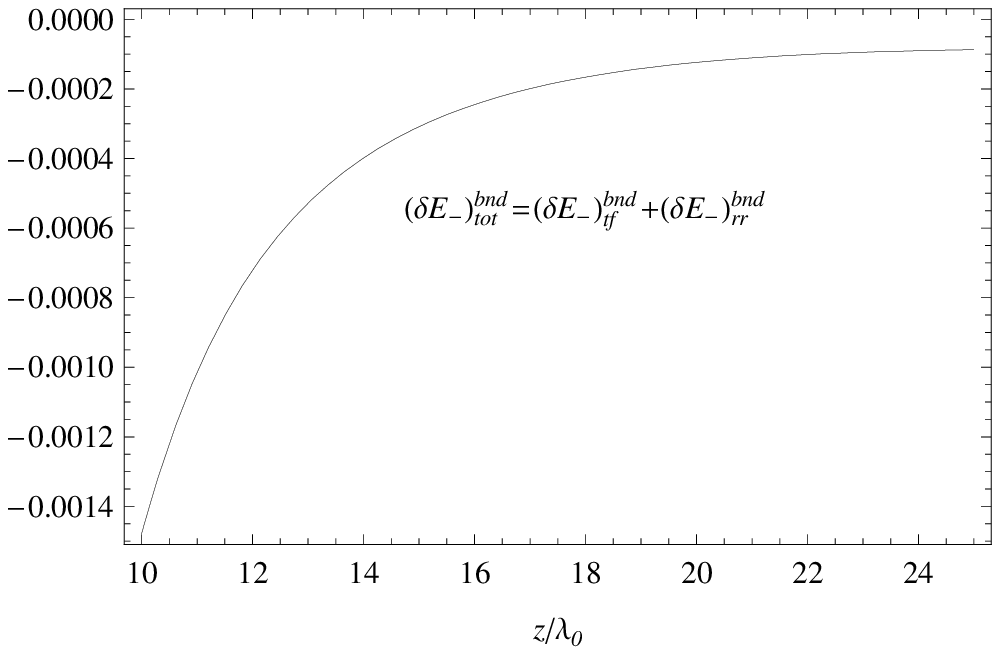}}
\subfigure[]{\label{fig:inte:e}
\includegraphics[height=2.3in,width=3.0in]{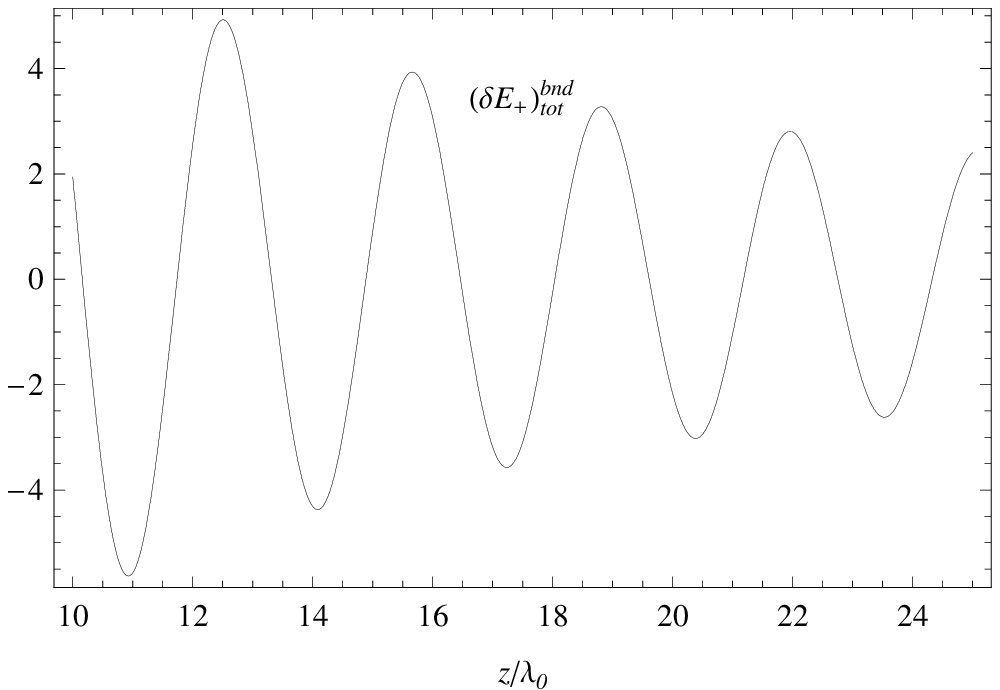}}
\caption{Energy level shifts of both a ground state atom and an
excited one polarized isotropically in the intermediate distance
regime ($\lambda_0\ll z\ll\beta$) at room temperature $T=300K$. Here
the frequency of Rb atom, $\omega_{0}=2.37\times10^{15}s^{-1}$ is
used, and the levels shifts are in the units of
${\alpha_0/(4\pi\varepsilon_0)}$. (a) The contributions of thermal
fluctuations (dash-dotted line) and radiation reaction (dashed line)
to the level shift of a ground state atom is plotted, demonstrating
the magnitude of two effects are comparable and there is no clear
domination of one over the other. (b) The level shift of a ground
state atom as a net result of two competing effects is plotted. The
net result is clearly much less than any of the competing effects.
(c) The energy level shift of an excited state atom,
Eq.~(\ref{intere2}), is plotted, showing clearly differences both in
magnitude and sign from that of a ground state. }\label{fig:inte:g}
\end{figure}

Now let us turn to intermediate distance regime, i.e., when
$\lambda_0\ll z\ll\beta$. Here one finds
\begin{eqnarray}
(\delta
E_-)^{bnd}_{tf}\approx-{\hbar\/4\pi\varepsilon_0}&\bigg[&\bigg({3\omega_0^3(\alpha_x+\alpha_y)\/8c^2z}
-{3\omega_0\/32z^3}(\alpha_x+\alpha_y+2\alpha_z)\bigg)\cos(2z\omega_0/c)
\nonumber\\&&-{3\omega_0^2\/16cz^2}(\alpha_x+\alpha_y+2\alpha_z)\sin(2z\omega_0/c)+{3c\/8\pi
z^4}(\alpha_x+\alpha_y+\alpha_z)\nonumber\\&&+{32\pi^5cz^2\/315\beta^6}(2\alpha_x+2\alpha_y-\alpha_z)\bigg]\;,
\end{eqnarray}
and
\begin{eqnarray}
(\delta
E_-)^{bnd}_{rr}\approx{\hbar\/4\pi\varepsilon_0}&\bigg[&\bigg({3\omega_0^3(\alpha_x+\alpha_y)\/8c^2z}
-{3\omega_0\/32z^3}(\alpha_x+\alpha_y+2\alpha_z)\bigg)\cos(2z\omega_0/c)\nonumber\\&&
-{3\omega_0^2\/16cz^2}(\alpha_x+\alpha_y+2\alpha_z)\sin(2z\omega_0/c)\bigg]\;.
\end{eqnarray}
Both the contributions of the thermal fluctuations and the radiation
reaction have terms which are oscillating functions of $z$. However,
these oscillating components cancel out when added up, leading to
\begin{eqnarray}
(\delta
E_-)^{bnd}_{tot}\approx-{\hbar\/4\pi\varepsilon_0}\bigg[{3c\/8\pi
z^4}(\alpha_x+\alpha_y+\alpha_z)+{32\pi^5cz^2\/315\beta^6}(2\alpha_x+2\alpha_y-\alpha_z)\bigg]\;,\label{interg}
\end{eqnarray}
where the first term is completely coincident with the zero
temperature Casimir-Polder interaction energy given in
Ref.~\cite{CP} and  the atom's polarization in different direction
contributes equally. The second  gives the thermal correction which
is equal to that in the short distance regime and it however depends
on the atom's polarization direction. One sees that the
Casimir-Polder force is a result of a residue energy shift due to
vacuum fluctuations after cancelations of competing oscillating
terms from thermal fluctuations and radiation reaction. This
demonstrates that the Casimir-Polder force originates from  a net
result of vacuum fluctuations at zero temperature and radiation
reaction. Here there is no clear domination of one effect over the
other as is in the short distance regime. In fact, the two effects
are now comparable and opposing each other, and the net effect is
much smaller than one either of them.  We demonstrate this feature
graphically in Fig.~\ref{fig:inte:tfrr} and Fig.~\ref{fig:inte:tot}.
For an isotropically polarized atom,  we obtain
\begin{eqnarray}
(\delta
E_-)^{bnd}_{tot}\approx-{\hbar\/4\pi\varepsilon_0}\bigg({3c\alpha_0\/8\pi
z^4} +{32\pi^5c\alpha_0z^2\/315\beta^6}\bigg)\;,\label{lowTinterg}
\end{eqnarray}
which is always negative  and the force on the ground state atom is
again attractive in the intermediate distance regime.

Finally, for the long distance regime, i.e., when
$z\gg\beta\gg\lambda_0$, one has
\begin{eqnarray}
(\delta
E_-)^{bnd}_{tf}\approx-{\hbar\/4\pi\varepsilon_0}&\bigg[&\bigg({3\omega_0^3(\alpha_x+\alpha_y)\/8c^2z}-
{3\omega_0\/32z^3}(\alpha_x+\alpha_y+2\alpha_z)\bigg)\cos(2z\omega_0/c)
\nonumber\\&&-{3\omega_0^2\/16cz^2}(\alpha_x+\alpha_y+2\alpha_z)\sin(2z\omega_0/c)
+{3c\/16z^3\beta}(\alpha_x+\alpha_y+2\alpha_z)\bigg]\;,\nonumber\\
\end{eqnarray}
and
\begin{eqnarray}
(\delta
E_-)^{bnd}_{rr}\approx{\hbar\/4\pi\varepsilon_0}&\bigg[&\bigg({3\omega_0^3(\alpha_x+\alpha_y)\/8c^2z}
-{3\omega_0\/32z^3}(\alpha_x+\alpha_y+2\alpha_z)\bigg)\cos(2z\omega_0/c)
\nonumber\\&&-{3\omega_0^2\/16cz^2}(\alpha_x+\alpha_y+2\alpha_z)\sin(2z\omega_0/c)\bigg]\;.
\end{eqnarray}
Similar to the intermediate distance regime ($\lambda_0\ll
z\ll\beta$), the oscillating terms in the contributions of thermal
fluctuations and radiation reaction balance each other. So, again,
in the present case, there is no clear domination of one effect over
the other. However, the position-dependent energy shift of the
ground state now has the form
\begin{eqnarray}
(\delta
E_-)^{bnd}_{tot}\approx-{\hbar\/4\pi\varepsilon_0}{3c\/16z^3\beta}(\alpha_x+\alpha_y+2\alpha_z)
=-{1\/4\pi\varepsilon_0}{3k_BT\/16z^3}(\alpha_x+\alpha_y+2\alpha_z)\;.\label{largezg}
\end{eqnarray}
This is linear in the temperature and depends on the polarization
direction of the atom too. In the present case, the contributions
from vacuum fluctuations are negligible as compared to that of
thermal fluctuations in clear contrast to the intermediate regime.
Therefore, in the low temperature limit ($\beta\gg\lambda_0$), only
when the distance between the ground state atom and the conducting
boundary is much larger than the thermal wavelength ($z\gg\beta$),
do the thermal fluctuations become the dominating source of  the
position-dependent energy shifts and thus of the force acting on the
atom. It is particularly interesting to note that the force acting
on a ground state atom which is transversely polarized is twice as
much as that of the atom which is  longitudinally polarized. For an
isotropically polarized ground state atom the total contributions
\begin{eqnarray}
(\delta
E_-)^{bnd}_{tot}\approx-{1\/4\pi\varepsilon_0}{\alpha_0k_BT\/4z^3}\;.\label{largezg2}
\end{eqnarray}
This is in agreement with the Lifshitz's result \cite{Lifshitz},
which was obtained in a macroscopic approach to the problem where
the force between two dielectric semi-infinite spaces at finite
temperature was considered and the force on an atom is obtained as a
limiting case when one  dielectric is sufficiently rarefied.  Note
that the force is again attractive.

\subsubsection{Excited state atom}

After having analyzed the force on an atom in the ground state in
the low temperature limit, let us now discuss what happens if the
atom is in the excited state. One notes, from
Eqs.~(\ref{inercpvf2e})-(\ref{inercprr2}), that the contributions of
thermal fluctuations to the energy shifts for the ground and excited
state atoms are of the same magnitude but opposite signs, while that
of the radiation reaction are equal. Since radiation reaction is
temperature independent, the  thermal corrections to energy shift of
the excited state come solely from thermal fluctuations and thus
will be just opposite to that of the ground state. As a result, in
the short distance regime where the contributions from radiation
reaction dominate over thermal fluctuations (and vacuum
fluctuations), one finds
\begin{eqnarray}
(\delta
E_+)^{bnd}_{tot}\approx-{\hbar\/4\pi\varepsilon_0}\bigg[{3\omega_{0}\/32z^3}(\alpha_x+\alpha_y+2\alpha_z)
-{32\pi^5cz^2\/315\beta^6}(2\alpha_x+2\alpha_y-\alpha_z)\bigg]\;,\label{lowTlowze}
\end{eqnarray}
which agrees with  the result for the ground state atom in the
leading order(refer to Eq.~(\ref{lowTlowz})). Therefore, in the low
temperature and short distance limit, the force acting on the atom
reduces to the Van der Waals result  for  both the ground and
excited in the leading order plus a small thermal corrections which
differs in sign. The net force is attractive. It is worth pointing
out here that the force on excited atoms, in general, may not be
necessarily the gradient of a potential.  However, in the
intermediate distance regime, where radiation reaction no longer
dominates, the total energy level shift deviates considerably from
that of the ground state as a result of the fact that the
oscillating terms from thermal fluctuations and radiation reaction
do not cancel as they do in the case of the ground state atom. The
result is
\begin{eqnarray}
(\delta
E_+)^{bnd}_{tot}\approx{\hbar\/4\pi\varepsilon_0}&\bigg[&\bigg({3\omega_0^3(\alpha_x+\alpha_y)\/4c^2z}
-{3\omega_0\/16z^3}(\alpha_x+\alpha_y+2\alpha_z)\bigg)\cos(2z\omega_0/c)
\nonumber\\&&-{3\omega_0^2\/8cz^2}(\alpha_x+\alpha_y+2\alpha_z)\sin(2z\omega_0/c)+{3c\/8\pi
z^4}(\alpha_x+\alpha_y+\alpha_z)\nonumber\\&&+{32\pi^5cz^2\/315\beta^6}(2\alpha_x+2\alpha_y-\alpha_z)\bigg]\;.\label{intere}
\end{eqnarray}
Amplitude of the oscillating terms is much larger than the
Casimir-Polder term which behaves like $1/z^4$ and consequently, for
a given atom, the position-dependent energy shift of the excited
state may be much larger than that of the ground state depending on
the distance $z$ of the atom from the boundary. The thermal
corrections are very small compared with the vacuum fluctuation
contributions. For an isotropically polarized atom, we can write
Eq.~(\ref{intere}) as
\begin{eqnarray}
(\delta
E_+)^{bnd}_{tot}\approx{\hbar\/4\pi\varepsilon_0}\bigg[\bigg({\omega_0^3\alpha_0\/2c^2z}
-{\omega_0\alpha_0\/4z^3}\bigg)\cos(2z\omega_0/c)
-{\omega_0^2\alpha_0\/2cz^2}\sin(2z\omega_0/c)+{3c\alpha_0\/8\pi
z^4}
+{32\pi^5c\alpha_0z^2\/315\beta^6}\bigg]\;,\label{intere2}\nonumber\\
\end{eqnarray}
which may be positive or negative depending on the distance $z$ so
that the force acting on the excited state atom may be attractive or
repulsive and even be zero. This is quite different from the case of
a ground state atom, where the position-dependent energy shift is
always negative and the force is attractive. It is interesting to
note that, for a given atom, there exist certain values of the
distance $z$ such that the oscillatory terms sum up to zero
%\begin{equation}
%({\omega_0^3\alpha_0\/2z}-{\omega_0\alpha_0\/8z^3})\cos(2z\omega_0)
%-{\omega_0^2\alpha_0\/2z^2}\sin(2z\omega_0)=0\;,{\label{Cond}}
%\end{equation}
and when that happens the position-dependent energy shift becomes
\begin{eqnarray}
(\delta
E_+)^{bnd}_{tot}\approx{\hbar\/4\pi\varepsilon_0}\bigg({3c\alpha_0\/8\pi
z^4} +{32\pi^5c\alpha_0z^2\/315\beta^6}\bigg)\;,
\end{eqnarray}
which leads to a force on the excited atom equal in magnitude to
that on ground state atom but opposite in direction. It is
remarkable that whenever the oscillatory terms do not sum up to
zero, the force on the excited atom is generally much stronger than
the usual Casimir-Polder force on the ground state atom and
alternates between attractive and repulsive as the distance varies.
The total energy shift for an excited state is plotted in
Fig.~\ref{fig:inte:e} along with Fig.~\ref{fig:inte:g} to show this.

Now it is time to look at the long distance regime. It is easy to
show that the energy shift can be approximated as
\begin{eqnarray}
(\delta
E_+)^{bnd}_{tot}\approx{\hbar\/4\pi\varepsilon_0}&\bigg[&\bigg({3\omega_0^3(\alpha_x+\alpha_y)\/4c^2z}
-{3\omega_0\/16z^3}(\alpha_x+\alpha_y+2\alpha_z)\bigg)\cos(2z\omega_0/c)
\nonumber\\&&-{3\omega_0^2\/8cz^2}(\alpha_x+\alpha_y+2\alpha_z)\sin(2z\omega_0/c)
+{3c\/16z^3\beta}(\alpha_x+\alpha_y+2\alpha_z)\bigg]\;.\label{largeze}
\end{eqnarray}
Here the oscillating terms arising from vacuum fluctuations are the
same as in the intermediate distance regime but the thermal
correction term ($1/z^3\beta$) overtakes the Casimir-Polder one
($1/z^4$) . A comparison with  Eq.~(\ref{largezg}) shows that,
although the thermal effects plays a major  role in the
position-dependent energy shift for the ground state, it is
generally not the case for the excited state because of the presence
of oscillating terms. In fact, the thermal correction term is of
higher order when compared with the oscillating terms, unless the
atom is such placed that the oscillatory terms sum up to zero.
Therefore, the net contributions of vacuum fluctuations and
radiation reaction in general dominate even in the long distance
regime if the atom is in the excited state, as opposed to the ground
state where thermal effects dominate in the long distance regime.
For an isotropically polarized atom,
\begin{eqnarray}
(\delta
E_+)^{bnd}_{tot}\approx{\hbar\/4\pi\varepsilon_0}\bigg[\bigg({\omega_0^3\alpha_0\/2c^2z}
-{\omega_0\alpha_0\/4z^3}\bigg)\cos(2z\omega_0/c)
-{\omega_0^2\alpha_0\/2cz^2}\sin(2z\omega_0/c)+{c\alpha_0\/4z^3\beta}\bigg]\;,\label{largeze2}
\end{eqnarray}
which again can be positive or negative depending on the position of
the atom.
%Behaviors  as a function of $z$ different from that of a
%ground state atom is shown in Fig.~\ref{fig:long}.

The above discussions reveal that the force acting on the excited
atom is definitely attractive only in the short distance regime and
can be either attractive or repulsive in the intermediate and long
distance regime. This is in a clear contrast to the ground state
atom where the force is attractive in all regimes. When the
oscillatory terms sum up to zero, we have
\begin{eqnarray}
(\delta
E_+)^{bnd}_{tot}\approx{\hbar\/4\pi\varepsilon_0}{c\alpha_0\/4z^3\beta}={1\/4\pi\varepsilon_0}{\alpha_0k_BT\/4z^3}\;,
\end{eqnarray}
leading to a force which is equal to the Lifshitz result in
magnitude but opposite in direction.

\subsection{High temperature limit}

Now we turn our attention to the high temperature limit, where we
assume  the wavelength of the thermal photons to be much smaller
than the transition wavelength of the atom, i.e.,
\begin{eqnarray}
\beta\ll\lambda_0\;,
\end{eqnarray}
which can also be written as $\beta\omega_0/c\ll1$. Then
Eqs.~(\ref{inercpvf2e}) and (\ref{inercpvf2g}), for the
contributions of the thermal fluctuations to the position-dependent
energy shifts, can be approximated as
\begin{eqnarray}
(\delta E_-)^{bnd}_{tf}=-(\delta
E_+)^{bnd}_{tf}\approx-{3\hbar\omega_{0}\alpha_j\/128\pi\varepsilon_0}\bigg[{2c\/\beta\omega_0}f_{j}(\omega_{0},z)
-g_{j}(\omega_{0},z,\beta)\bigg]\;.\label{highTeg}
\end{eqnarray}
As in the low temperature limit, we will separately examine the
position-dependent energy  shifts and thus the force acting on the
atom for both the ground and excited states in three different
distance regimes.

 We first consider the case when the atom is placed so close to the boundary that the
distance $z$ is much smaller than the thermal wavelength
($z\ll\beta\ll\lambda_0$). It  follows that contributions of thermal
fluctuations and radiation reaction to the energy level shifts are
now respectively,
\begin{eqnarray}
(\delta E_-)^{bnd}_{tf}=-(\delta E_+)^{bnd}_{tf}\approx
{\hbar\/4\pi\varepsilon_0}{3c\/16\beta
z^3}(\alpha_x+\alpha_y+2\alpha_z)\;,\label{highTshortztf}
\end{eqnarray}and
\begin{eqnarray}
(\delta E_-)^{bnd}_{rr}=(\delta E_+)^{bnd}_{rr}\approx
-{\hbar\/4\pi\varepsilon_0}{3\omega_0\/32z^3}(\alpha_x+\alpha_y+2\alpha_z)\;.\label{highTshortzrr}
\end{eqnarray}
Here the contributions of thermal fluctuations (\ref{highTshortztf})
are much larger than that of the radiation reaction
(\ref{highTshortzrr}) and play the dominating role in the
position-dependent energy shifts for both the ground and excited
state. It is interesting to note that in the high temperature limit,
even when the distance $z$ is much smaller than the wavelength of
thermal photons, the thermal effects dominate over zero-point
contributions. This is in contrast to the low temperature limit
where thermal effects dominate only in the long distance regime
when the distance $z$ is much larger than the wavelength of thermal
photons.
  For an isotropically
polarized atom, we obtain
\begin{eqnarray}
(\delta E_-)^{bnd}_{tot}=-(\delta
E_+)^{bnd}_{tot}\approx{\hbar\/4\pi\varepsilon_0}{c\alpha_0\/4\beta
z^3}={1\/4\pi\varepsilon_0}{\alpha_0k_BT\/4z^3}\;,
\end{eqnarray}
which is linear in  temperature, and gives a positive
position-dependent energy shift for a ground state atom and a
negative one for an excited state. This means that the force acting
on an atom is repulsive if the atom is in the ground state and
attractive if otherwise. Notice that the force on a ground state
atom in the present case is different from the Lifshitz's result in
the low temperature and long distance limit in direction although it
is equal in magnitude.

If  the distance $z$ is much larger than the thermal wavelength but
much smaller than the transition wavelength, i.e., $\beta\ll
z\ll\lambda_0$, then we find
\begin{eqnarray}
(\delta E_-)^{bnd}_{tf}=-(\delta
E_+)^{bnd}_{tf}\approx-{\hbar\/4\pi\varepsilon_0}\bigg[
{3\omega_0^2\/8c\beta z}(\alpha_x+\alpha_y-2\alpha_z)
-{9\omega_0^4z\/8c^3\beta}\bigg(\alpha_x+\alpha_y-{2\/3}\alpha_z\bigg)\bigg]\;,
\end{eqnarray}
and
\begin{eqnarray}
(\delta E_-)^{bnd}_{rr}=(\delta E_+)^{bnd}_{rr}\approx
-{\hbar\/4\pi\varepsilon_0}{3\omega_0\/32z^3}(\alpha_x+\alpha_y+2\alpha_z)\;.
\end{eqnarray}
For the total position-dependent energy shifts, we can obtain
\begin{eqnarray}
(\delta E_-)^{bnd}_{tot}\approx
-{\hbar\/4\pi\varepsilon_0}\bigg[{3\omega_0^2\/8c\beta
z}(\alpha_x+\alpha_y-2\alpha_z)
-{9\omega_0^4z\/8c^3\beta}\bigg(\alpha_x+\alpha_y-{2\/3}\alpha_z\bigg)
+{3\omega_0\/32z^3}(\alpha_x+\alpha_y+2\alpha_z)\bigg]\;,\label{highTinterzg}\nonumber\\
\end{eqnarray}
for the ground state, and
\begin{eqnarray}
(\delta E_+)^{bnd}_{tot}\approx
{\hbar\/4\pi\varepsilon_0}\bigg[{3\omega_0^2\/8c\beta
z}(\alpha_x+\alpha_y-2\alpha_z)
-{9\omega_0^4z\/8c^3\beta}\bigg(\alpha_x+\alpha_y-{2\/3}\alpha_z\bigg)
-{3\omega_0\/32z^3}(\alpha_x+\alpha_y+2\alpha_z)\bigg]\;.\label{highTinterze}\nonumber\\
\end{eqnarray}
for the excited state. Here there is again no strong domination of
one effect over the other. However, the above results show that now
whether the force acting on the atom is attractive or repulsive
depends crucially on the polarizations of the atom. For an example,
a change of relative ratio of polarizations in different directions
may result in a change of direction of the force.
 We note that for an isotropically
polarized atom, the first term in both Eq.(\ref{highTinterzg}) and
(\ref{highTinterze}) is zero, so
\begin{eqnarray}
(\delta E_-)^{bnd}_{tot}\approx{\hbar\/4\pi\varepsilon_0}\bigg(
{\omega_0^4\alpha_0z\/2c^3\beta}-{\omega_0\alpha_0\/8z^3}\bigg)\;,
\end{eqnarray}
and
\begin{eqnarray}
(\delta E_+)^{bnd}_{tot}\approx-{\hbar\/4\pi\varepsilon_0}\bigg(
{\omega_0^4\alpha_0z\/2c^3\beta}+{\omega_0\alpha_0\/8z^3}\bigg)\;.
\end{eqnarray}
Here for an excited state atom, the position-dependent energy shift
is always negative and induces an attractive force, but for a ground
state atom, it is determined by the magnitude of the distance $z$
and the wavelength of the thermal photon. The crucial dependence of
the direction of the  force acting the atom on its polarization in
this case is an interesting feature of atom-wall interactions
revealed in the present paper.

Thirdly, for the long distance limit such that the distance $z$ is
much bigger than the transition wavelength
($z\gg\lambda_0\gg\beta$), the contributions of thermal fluctuations
and radiation reaction can be written as
\begin{eqnarray}
(\delta E_-)^{bnd}_{tf}=-(\delta E_+)^{bnd}_{tf}&\approx
&-{\hbar\/4\pi\varepsilon_0}\bigg[\bigg({3\omega_0^2(\alpha_x+\alpha_y)\/4c\beta
z}-{3c\/16\beta
z^3}(\alpha_x+\alpha_y+2\alpha_z)\bigg)\cos(2z\omega_0/c)\nonumber\\&&\quad\quad\quad\-{3\omega_0\/8\beta
z^2}(\alpha_x+\alpha_y+2\alpha_z)\sin(2z\omega_0/c)+{3c\/16\beta
z^3}(\alpha_x+\alpha_y+2\alpha_z)\bigg]\;,\nonumber\\
\end{eqnarray}and
\begin{eqnarray}
(\delta E_-)^{bnd}_{rr}=(\delta
E_+)^{bnd}_{rr}\approx{\hbar\/4\pi\varepsilon_0}&\bigg[&\bigg({3\omega_0^3(\alpha_x+\alpha_y)\/8c^2z}
-{3\omega_0\/32z^3}(\alpha_x+\alpha_y+2\alpha_z)\bigg)\cos(2z\omega_0/c)\nonumber\\&&
-{3\omega_0^2\/16cz^2}(\alpha_x+\alpha_y+2\alpha_z)\sin(2z\omega_0/c)\bigg]\;.
\end{eqnarray}
Clearly, in the present case,  the contribution of radiation
reaction is much smaller than that of thermal fluctuations since
$\beta\omega_0/c\ll 1$ and thus the vacuum fluctuation effect
dominates. So the position-dependent energy shifts of both the
ground and excited states are dominated by the contribution of
thermal fluctuations. Assuming isotropic polarization, we arrive at
\begin{eqnarray}
(\delta E_-)^{bnd}_{tot}=-(\delta
E_+)^{bnd}_{tot}&\approx&-{\hbar\/4\pi\varepsilon_0}\bigg[\bigg({\omega_0^2\alpha_0\/2c\beta
z}-{c\alpha_0\/4\beta
z^3}\bigg)\cos(2z\omega_0/c)-{\omega_0\alpha_0\/2\beta
z^2}\;\sin(2z\omega_0/c)+{c\alpha_0\/4\beta
z^3}\bigg]\nonumber\\&\approx&-{\hbar\/4\pi\varepsilon_0}\bigg[{\omega_0^2\alpha_0\/2c\beta
z}\cos(2z\omega_0/c)-{\omega_0\alpha_0\/2\beta
z^2}\;\sin(2z\omega_0/c)+{c\alpha_0\/4\beta z^3}\bigg]\;,\nonumber\\
\end{eqnarray}
which has both terms of oscillating function of  distance $z$ and a
Lifshitz-like term.  If the atom is so placed that the oscillating
part is equal to zero, the position-dependent energy shift of the
ground state reduces to the Lifshitz's result.

In all the discussions above, we have only studied energy level
shifts of an atom in its ground and excited state separately. An
interesting issue yet to be addressed is the thermal average of
energy level shifts of an atom in the equilibrium with thermal
photons. This is particularly desirable in the high temperature
limit, since now  the ground state atoms have a high possibility to
absorb thermal photons to transit to the excited state.  Taking into
account the fermionic nature of the two-level atom, which can be
seen by noting that  the atomic raising and lowering operators obey
the anticommutation relation, $\{S_+,S_-\}=1$, we can calculate the
thermal average of contributions of the thermal fluctuations and
radiation reaction to the position-dependent energy shifts as
follows,
\begin{eqnarray}
({\overline{\delta
E}})^{bnd}_{tf}&=&{1\/1+e^{-\omega_{0}\beta/c}}(\delta
E_-)^{bnd}_{tf} +\bigg(1-{1\/1+e^{-\omega_{0}\beta/c}}\bigg)(\delta
E_+)^{bnd}_{tf}
\nonumber\\&=&{3\hbar\omega_{0}\alpha_j\/128\pi\varepsilon_0}~
\bigg[-f_j(\omega_{0},z)+{1-e^{-\omega_{0}\beta/c}\/1+e^{-\omega_{0}\beta/c}}~g_j(\omega_{0},z,\beta)\bigg]\;,\label{averagetf}
\end{eqnarray}
and
\begin{eqnarray}
({\overline{\delta
E}})^{bnd}_{rr}&=&{1\/1+e^{-\omega_{0}\beta/c}}(\delta
E_-)^{bnd}_{rr} +\bigg(1-{1\/1+e^{-\omega_{0}\beta/c}}\bigg)(\delta
E_+)^{bnd}_{rr}
\nonumber\\&=&{3\hbar\omega_{0}\alpha_j\/128\pi\varepsilon_0}~
f_j(\omega_{0},z)\;.\label{averagerr}
\end{eqnarray}
When the temperature is low, the thermal average shift is
essentially the shift of the ground state as expected, and
significant deviations only occur when  temperature is high.  So, in
what follows, we will examine in detail the average energy shifts in
the three distinct distance regimes in the high temperature limit.

First, the short distance regime, where  the distance $z$ is assumed
to be much smaller than the thermal wavelength, i.e.,
$z\ll\beta\ll\lambda_0$. Now the contributions of thermal
fluctuations and radiation reaction to the average energy shifts can
be approximated as
\begin{eqnarray}
({\overline{\delta
E}})^{bnd}_{tf}\approx{\hbar\/4\pi\varepsilon_0}\bigg({3\omega_0\/32z^3}-{3\omega_{0}^2\beta\/64c
z^3}\bigg)(\alpha_x+\alpha_y+2\alpha_z)\;,\label{highTlowztf}
\end{eqnarray}and
\begin{eqnarray}
({\overline{\delta
E}})^{bnd}_{rr}\approx-{\hbar\/4\pi\varepsilon_0}{3\omega_0\/32z^3}(\alpha_x+\alpha_y+2\alpha_z)\;.\label{highTlowzrr}
\end{eqnarray}
Here one can see that  the contribution of temperature-independent
vacuum fluctuations to the average energy shift cancels  that of the
radiation reaction when added up, leaving a correction due to
thermal fluctuations as the dominant term in the average energy
shift. So
\begin{eqnarray}
({\overline{\delta
E}})^{bnd}_{tot}\approx-{\hbar\/4\pi\varepsilon_0}{3\omega_{0}^2\beta\/64c
z^3}(\alpha_x+\alpha_y+2\alpha_z)=-{1\/4\pi\varepsilon_0}{3\hbar^2\omega_{0}^2\/64
k_BTz^3}(\alpha_x+\alpha_y+2\alpha_z)\;,\label{highTlowz}
\end{eqnarray}
This means that the average energy shift decreases as the
temperature increases in the short distance regime
($z\ll\beta\ll\lambda_0$). For isotropic polarization,
Eq.~(\ref{highTlowz}) becomes
\begin{eqnarray}
({\overline{\delta
E}})^{bnd}_{tot}\approx-{1\/4\pi\varepsilon_0}{\hbar^2\omega_{0}^2\alpha_0\/16
k_BTz^3}\;.
\end{eqnarray}
This is always negative and leads to an attractive average force on
the atom.

Second,  we consider the intermediate distance regime, where the
distance $z$ is much larger than the thermal wavelength but much
smaller than the transition wavelength of the atom, i.e., $\beta\ll
z\ll\lambda_0$. Now one finds
\begin{eqnarray}
({\overline{\delta
E}})^{bnd}_{tf}\approx-{\hbar\/4\pi\varepsilon_0}\bigg[{3\omega_{0}^3\/16c^2z}(\alpha_x+\alpha_y-2\alpha_z)
+{3\omega_{0}^2\/4\pi c
z^2}\alpha_z-{\omega_{0}^3\beta^2\/128c^2z^3}(\alpha_x+\alpha_y+2\alpha_z)\bigg]\;,
\end{eqnarray}
\begin{eqnarray}
({\overline{\delta
E}})^{bnd}_{rr}\approx-{\hbar\/4\pi\varepsilon_0}{3\omega_{0}\/32z^3}
(\alpha_x+\alpha_y+2\alpha_z)\;,
\end{eqnarray}
so the average energy shift is given by
\begin{eqnarray}
({\overline{\delta
E}})^{bnd}_{tot}\approx-{\hbar\/4\pi\varepsilon_0}\bigg({3\omega_{0}\/32z^3}-{\omega_{0}^3\beta^2\/128c^2z^3}\bigg)
(\alpha_x+\alpha_y+2\alpha_z)\;.\label{highT}
\end{eqnarray}
Here the average energy shift is caused mainly by the radiation
reaction and the thermal correction due to thermal fluctuations is
quadratic in  $\beta$ in contrast to the linear dependence in the
short distance regime. For an isotropically polarized atom,
Eq.~(\ref{highT}) reduces to
\begin{eqnarray}
({\overline{\delta
E}})^{bnd}_{tot}\approx-{\hbar\/4\pi\varepsilon_0}\bigg({\omega_0\alpha_0\/8z^3}
-{\omega_{0}^3\beta^2\alpha_0\/96c^2z^3}\bigg)\;,\label{highT2}
\end{eqnarray}
where the first term is just the Van der Waals result.

Finally,  we can estimate the contributions of thermal fluctuations
and radiation reaction to the average energy shift, in the long
distance regime ($z\gg\lambda_0\gg\beta$), to get
\begin{eqnarray}
({\overline{\delta
E}})^{bnd}_{tf}\approx-{\hbar\/4\pi\varepsilon_0}&\bigg[&\bigg({3\omega_0^3(\alpha_x+\alpha_y)\/8c^2z}
-{3\omega_0\/32z^3}(\alpha_x+\alpha_y+2\alpha_z)\bigg)\cos(2z\omega_0/c)\nonumber\\&&
-{3\omega_0^2\/16cz^2}(\alpha_x+\alpha_y+2\alpha_z)\sin(2z\omega_0/c)
+\bigg({3\omega_0\/32z^3}-{\omega_0^3\beta^2\/128c^2z^3}\bigg)(\alpha_x+\alpha_y+2\alpha_z)\bigg]\;,\label{highTlargeztf}\nonumber\\
\end{eqnarray}and
\begin{eqnarray}
({\overline{\delta
E}})^{bnd}_{rr}\approx{\hbar\/4\pi\varepsilon_0}&\bigg[&\bigg({3\omega_0^3(\alpha_x+\alpha_y)\/8c^2z}
-{3\omega_0\/32z^3}(\alpha_x+\alpha_y+2\alpha_z)\bigg)\cos(2z\omega_0/c)\nonumber\\&&
-{3\omega_0^2\/16cz^2}(\alpha_x+\alpha_y+2\alpha_z)\sin(2z\omega_0/c)
\bigg]\;.\label{highTlargezrr}
\end{eqnarray}
Noting that the oscillating terms in the contributions of thermal
fluctuations are canceled out by that of the radiation reaction in
the calculation of the total shift,  we obtain
\begin{eqnarray}
({\overline{\delta
E}})^{bnd}_{tot}\approx-{\hbar\/4\pi\varepsilon_0}\bigg({3\omega_{0}\/32z^3}-{\omega_{0}^3\beta^2\/128c^2z^3}\bigg)
(\alpha_x+\alpha_y+2\alpha_z)\;
\end{eqnarray}
This coincides with Eq.~(\ref{highT}) for the case of $\beta\ll
z\ll\lambda_0$ in mathematical appearance but differs in physical
origin. The reason is that here the average energy shift comes
mainly from the contribution of the thermal fluctuations while in
the intermediate distance regime the radiation reaction plays the
major role.

\section{Conclusion}

Using the DDC formalism which separates the contributions of thermal
fluctuations  and radiation reaction and allows a distinct
microscopic treatment to atoms in the ground and excited states, in
contrast to the macroscopic approach where atoms are treated as a
limiting case of a dielectric, we have investigated the modification
by the presence of a plane wall of energy level shifts of two-level
atoms which are in multipolar coupling with quantized
electromagnetic fields at the finite temperature. The position
dependent energy shifts give rise to an induced force acting on the
atoms. We have examined the behaviors of this force in both the low
temperature limit and the high temperature limit for both the ground
state and excited state atoms.

In the low temperature limit where the wavelength of thermal photons
is assumed to be much larger than the transition wavelength of the
atom,  our calculations reproduce, for a ground state atom, the van
der Waals force at short distance regime, the Casmir-Polder force in
the intermediate regime and the Lifshitz result in the long distance
regime,  and show that the force is attractive in all three regimes.
However if the atom is in the excited state, the atom-wall force
agrees with that in the ground state in the leading order only in
the short distance limit. In the intermediate distance regime the
force on the excited state atoms is generally strong than the
Casimir-Polder force on the ground state atoms and  alternates
between attractive and repulsive as the atom's distance from the
wall  varies. In the long distance regime, the main source of the
force is the net contributions of competing vacuum fluctuation and
radiation reaction and the thermal correction term is in general of
higher order. This is in clear contrast to the ground state atom
where the thermal effect (the Lifshitz term) dominates. Furthermore,
the force can again be attractive and repulsive depending on the
atom's distance.

In the high temperature limit  where the wavelength of thermal
photons is assumed to be much smaller than the transition wavelength
of the atom,  we find that the thermal effects dominate over
zero-point contributions even in the short distance limit where the
distance is small as compared to the transition wavelength and the
force is linear in temperature in the leading order in all three
regimes. In the short distance regime the force on an excited atom
is attractive while that on a ground state atom is repulsive. In the
intermediate distance, the direction of the force acting on the
atoms depends crucially the polarization. Only in the long distance
regime does the force have oscillatory part, leading to a force
which would either be attractive or repulsive.

We also have calculated the thermal average of energy level shifts
for an ensemble of atoms in equilibrium with the thermal photons in
the high temperature limit. We find that the average force is
attractive and decreases as $1/T$ as temperature $T$ grows in the
short distance regime. However, both in the intermediate and long
distance regimes, the average force coincides with the van de Waals
force  in the leading order, which is curiously independent of the
temperature. One should note, however, that although the force has
the same form in two different regimes, the physical origin is quite
different, with the force in the intermediate regime being generated
by  radiation reaction and that in long distance by thermal
fluctuations.

By not treating atoms as a limiting case of a macroscopic
dielectric, we are able to reveal the influence of different atomic
polarizations on the energy level shifts and thus the force acting
on the atoms. Interestingly, we find that  both the magnitude and
the direction of the force acting on an atom have a clear dependence
on the atomic polarization directions. In certain cases, a change of
relative ratio of polarizations in different directions may result
in a change of direction of the force.

Finally, as far as the separation of effects of fluctuations and
radiation reaction in the DDC prescription itself is concerned, we
only find a clear domination of one over the other in short distance
regime in the low temperature limit where radiation reaction effect
dominates and in both the short and long distance regimes in the
high temperature limit where fluctuations effect does. An
interesting issue for future discussions is the non-equilibrium
situation where the atom and field have different temperatures and
the two terms can thus  be varied independently. In fact, in this
case, the contribution of radiation reaction (atom fluctuations)
would only depend on the atomic temperature and the thermal
fluctuation contribution (field fluctuations) would only depend on
the field temperature. We hope to return to this issue in the
future.

\begin{acknowledgments}
This work was supported in part by the National Natural Science
Foundation of China under Grants No. 10575035, 10775050, the SRFDP
under Grant No. 20070542002, and the Hunan Provincial Innovation
Foundation for Postgraduate.
\end{acknowledgments}

\end{document}